# Testing Einstein's special relativity with Fermi's short hard γ-ray burst GRB090510

A. A. Abdo[2,3], M. Ackermann[4], M. Ajello[4], K. Asano[5,6], W. B. Atwood[7], M. Axelsson[8,9], L. Baldini[10], J. Ballet[11], G. Barbiellini[12,13], M. G. Baring[14], D. Bastieri[15,16], K. Bechtol[4], R. Bellazzini[10], B. Berenji[4], P. N. Bhat[17], E. Bissaldi[18], E. D. Bloom[4], E. Bonamente[19,20], J. Bonnell[21], A. W. Borgland[4], J. Bregeon[10], A. Brez[10], M. S. Briggs[17], M. Brigida[22,23], P. Bruel[24], T. H. Burnett[25], G. A. Caliandro[22,23], R. A. Cameron[4], P. A. Caraveo[26], J. M. Casandjian[11], C. Cecchi[19,20], Ö. Çelik[21], V. Chaplin[17], E. Charles[4], C. C. Cheung[21], J. Chiang[4], S. Ciprini[19,20], R. Claus[4], J. Cohen-Tanugi[27], L. R. Cominsky[28], V. Connaughton[17], J. Conrad[29,9,30,31], S. Cutini[32], C. D. Dermer[2], A. de Angelis[33], F. de Palma[22,23], S. W. Digel[4], B. L. Dingus[34], G. Di Bernardo[10], E. do Couto e Silva[4], P. S. Drell[4], R. Dubois[4], D. Dumora[35,36], C. Farnier[27], C. Favuzzi[22,23], S. J. Fegan[24], J. Finke[2,3], G. Fishman[37], W. B. Focke[4], L. Foschini[38], Y. Fukazawa[39], S. Funk[4], P. Fusco[22,23], F. Gargano[23], D. Gasparrini[32], N. Gehrels[21,40], L. Gibby[41], B. Giebels[24], N. Giglietto[22,23], F. Giordano[22,23], T. Glanzman[4], G. Godfrey[4], J. Granot[42,1], J. Greiner[18], I. A. Grenier[11], M.-H. Grondin[35,36], J. E. Grove[2], D. Grupe[43], L. Guillemot[35,36], S. Guiriec[17,1], Y. Hanabata[39], A. K. Harding[21], M. Hayashida[4], E. Hays[21], E. Hoversten[43], R. E. Hughes[44], G. Jóhannesson[4], A. S. Johnson[4], R. P. Johnson[7], W. N. Johnson[2], T. Kamae[4], H. Katagiri[39], J. Kataoka[5,45], N. Kawai[5,46], M. Kerr[25], R. M. Kippen[34], J. Knödlseder[47], D. Kocevski[4], C. Kouveliotou[37], F. Kuehn[44], M. Kuss[10], J. Lande[4], L. Latronico[10], M. Lemoine-Goumard[35,36], F. Longo[12,13], F. Loparco[22,23], B. Lott[35,36], M. N. Lovellette[2], P. Lubrano[19,20], G. M. Madejski[4], A. Makeev[2,48], M. N. Mazziotta[23],

S. McBreen[18,49], J. E. McEnery[21], S. McGlynn[30,9], P. Mészáros[43], C. Meurer[29,9], P. F. Michelson[4], W. Mitthumsiri[4], T. Mizuno[39], A. A. Moiseev[50,40], C. Monte[22,23], M. E. Monzani[4], E. Moretti[51,12,13], A. Morselli[52], I. V. Moskalenko[4], S. Murgia[4], P. L. Nolan[4], J. P. Norris[53], E. Nuss[27], M. Ohno[54,1], T. Ohsugi[39], N. Omodei[10], E. Orlando[18], J. F. Ormes[53], M. Ozaki[54], W. S. Paciesas[17], D. Paneque[4], J. H. Panetta[4], D. Parent[35,36], V. Pelassa[27,1], M. Pepe[19,20], M. Pesce-Rollins[10], V. Petrosian[4], F. Piron[27], T. A. Porter[7], R. Preece[17], S. Rainò[22,23], E. Ramirez-Ruiz[55], R. Rando[15,16], M. Razzano[10], S. Razzaque[2,3], A. Reimer[56,4], O. Reimer[56,4], T. Reposeur[35,36], S. Ritz[7], L. S. Rochester[4], A. Y. Rodriguez[57], M. Roth[25], F. Ryde[30,9], H. F.-W. Sadrozinski[7], D. Sanchez[24], A. Sander[44], P. M. Saz Parkinson[7], J. D. Scargle[58], T. L. Schalk[7], C. Sgrò[10], E. J. Siskind[59], D. A. Smith[35,36], P. D. Smith[44], G. Spandre[10], P. Spinelli[22,23], M. Stamatikos[21], F. W. Stecker[21], M. S. Strickman[2], D. J. Suson[60], H. Tajima[4], H. Takahashi[39], T. Takahashi[54], T. Tanaka[4], J. B. Thayer[4], J. G. Thayer[4], D. J. Thompson[21], L. Tibaldo[15,11,16], K. Toma[43], D. F. Torres[61,57], G. Tosti[19,20], E. Troja[5], Y. Uchiyama[54,4], T. Uehara[39], T. L. Usher[4], A. J. van der Horst[37,62], V. Vasileiou[21,50,63], N. Vilchez[47], V. Vitale[52,64], A. von Kienlin[18], A. P. Waite[4], P. Wang[4], C. Wilson-Hodge[37], K. S. Wood[2], X. F. Wu[43,65,66], R. Yamazaki[39], T. Ylinen[30,67,9], M. Ziegler[7]

[1]Corresponding authors: J. Granot, j.granot@herts.ac.uk ; S. Guiriec, sylvain.guiriec@nasa.gov ; M. Ohno, ohno@astro.isas.jaxa.jp; V. Pelassa, pelassa@lpta.in2p3.fr.

[2]Space Science Division, Naval Research Laboratory, Washington, DC 20375, USA

[3]National Research Council Research Associate, National Academy of Sciences, Washington, DC 20001, USA

[4]W. W. Hansen Experimental Physics Laboratory, Kavli Institute for Particle Astrophysics and Cosmology, Department of Physics and SLAC National Accelerator Laboratory, Stanford University, Stanford, CA 94305, USA

[5]Department of Physics, Tokyo Institute of Technology, Meguro City, Tokyo 152-8551, Japan

[6]Interactive Research Center of Science, Tokyo Institute of Technology, Meguro City, Tokyo 152-8551, Japan

[7]Santa Cruz Institute for Particle Physics, Department of Physics and Department of Astronomy and Astrophysics, University of California at Santa Cruz, Santa Cruz, CA 95064, USA

[8]Department of Astronomy, Stockholm University, SE-106 91 Stockholm, Sweden

[9]The Oskar Klein Centre for Cosmo Particle Physics, AlbaNova, SE-106 91 Stockholm, Sweden

[10]Istituto Nazionale di Fisica Nucleare, Sezione di Pisa, I-56127 Pisa, Italy

[11]Laboratoire AIM, CEA-IRFU/CNRS/Université Paris Diderot, Service d'Astrophysique, CEA Saclay, 91191 Gif sur Yvette, France

[12]Istituto Nazionale di Fisica Nucleare, Sezione di Trieste, I-34127 Trieste, Italy

[13]Dipartimento di Fisica, Università di Trieste, I-34127 Trieste, Italy

[14]Rice University, Department of Physics and Astronomy, MS-108, P. O. Box 1892, Houston, TX 77251, USA

[15]Istituto Nazionale di Fisica Nucleare, Sezione di Padova, I-35131 Padova, Italy

[16]Dipartimento di Fisica “G. Galilei”, Università di Padova, I-35131 Padova, Italy

[17]University of Alabama in Huntsville, Huntsville, AL 35899, USA

[18]Max-Planck Institut für extraterrestrische Physik, 85748 Garching, Germany

[19]Istituto Nazionale di Fisica Nucleare, Sezione di Perugia, I-06123 Perugia, Italy

[20]Dipartimento di Fisica, Università degli Studi di Perugia, I-06123 Perugia, Italy

[21]NASA Goddard Space Flight Center, Greenbelt, MD 20771, USA

[22]Dipartimento di Fisica "M. Merlin" dell'Università e del Politecnico di Bari, I-70126 Bari, Italy

[23]Istituto Nazionale di Fisica Nucleare, Sezione di Bari, 70126 Bari, Italy

[24]Laboratoire Leprince-Ringuet, École polytechnique, CNRS/IN2P3, Palaiseau, France

[25]Department of Physics, University of Washington, Seattle, WA 98195-1560, USA

[26]INAF-Istituto di Astrofisica Spaziale e Fisica Cosmica, I-20133 Milano, Italy

[27]Laboratoire de Physique Théorique et Astroparticules, Université Montpellier 2, CNRS/IN2P3, Montpellier, France

[28]Department of Physics and Astronomy, Sonoma State University, Rohnert Park, CA 94928-3609, USA

[29]Department of Physics, Stockholm University, AlbaNova, SE-106 91 Stockholm, Sweden

[30]Department of Physics, Royal Institute of Technology (KTH), AlbaNova, SE-106 91 Stockholm, Sweden

[31]Royal Swedish Academy of Sciences Research Fellow, funded by a grant from the K. A. Wallenberg Foundation

[32]Agenzia Spaziale Italiana (ASI) Science Data Center, I-00044 Frascati (Roma), Italy

[33]Dipartimento di Fisica, Università di Udine and Istituto Nazionale di Fisica Nucleare, Sezione di Trieste, Gruppo Collegato di Udine, I-33100 Udine, Italy

[34]Los Alamos National Laboratory, Los Alamos, NM 87545, USA

[35]Université de Bordeaux, Centre d'Études Nucléaires Bordeaux Gradignan, UMR 5797, Gradignan, 33175, France

[36]CNRS/IN2P3, Centre d'Études Nucléaires Bordeaux Gradignan, UMR 5797, Gradignan, 33175, France

[37]NASA Marshall Space Flight Center, Huntsville, AL 35812, USA

[38]INAF Osservatorio Astronomico di Brera, I-23807 Merate, Italy

[39]Department of Physical Sciences, Hiroshima University, Higashi-Hiroshima, Hiroshima 739-8526, Japan

[40]University of Maryland, College Park, MD 20742, USA

[41]Science Applications International Corporation, Huntsville, AL 35899, USA

[42]Centre for Astrophysics Research, University of Hertfordshire, College Lane, Hatfield AL10 9AB , UK

[43]Department of Astronomy and Astrophysics, Pennsylvania State University, University Park, PA 16802, USA

[44]Department of Physics, Center for Cosmology and Astro-Particle Physics, The Ohio State University, Columbus, OH 43210, USA

[45]Waseda University, 1-104 Totsukamachi, Shinjuku-ku, Tokyo, 169-8050, Japan

[46]Cosmic Radiation Laboratory, Institute of Physical and Chemical Research (RIKEN), Wako, Saitama 351-0198, Japan

[47]Centre d'Étude Spatiale des Rayonnements, CNRS/UPS, BP 44346, F-30128 Toulouse Cedex 4, France

[48]George Mason University, Fairfax, VA 22030, USA

[49]University College Dublin, Belfield, Dublin 4, Ireland

[50]Center for Research and Exploration in Space Science and Technology (CRESST), NASA Goddard Space Flight Center, Greenbelt, MD 20771, USA

[51]Istituto Nazionale di Fisica Nucleare, Sezione di Trieste, and Università di Trieste, I-34127 Trieste, Italy

[52]Istituto Nazionale di Fisica Nucleare, Sezione di Roma "Tor Vergata", I-00133 Roma, Italy

[53]Department of Physics and Astronomy, University of Denver, Denver, CO 80208, USA

[54]Institute of Space and Astronautical Science, JAXA, 3-1-1 Yoshinodai, Sagamihara, Kanagawa 229-8510, Japan

[55]UCO/Lick Observatories, Santa Cruz, CA 95064, USA

[56]Institut für Astro- und Teilchenphysik and Institut für Theoretische Physik, Leopold-Franzens-Universität Innsbruck, A-6020 Innsbruck, Austria

[57]Institut de Ciencies de l'Espai (IEEC-CSIC), Campus UAB, 08193 Barcelona, Spain

[58]Space Sciences Division, NASA Ames Research Center, Moffett Field, CA 94035-1000, USA

[59]NYCB Real-Time Computing Inc., Lattingtown, NY 11560-1025, USA

[60]Department of Chemistry and Physics, Purdue University Calumet, Hammond, IN 46323-2094, USA

[61]Institució Catalana de Recerca i Estudis Avançats, Barcelona, Spain

[62]NASA Postdoctoral Program Fellow, USA

[63]University of Maryland, Baltimore County, Baltimore, MD 21250, USA

[64]Dipartimento di Fisica, Università di Roma "Tor Vergata", I-00133 Roma, Italy

[65]Joint Center for Particle Nuclear Physics and Cosmology (J-CPNPC), Nanjing

210093, China

[66]Purple Mountain Observatory, Chinese Academy of Sciences, Nanjing 210008, China

[67]School of Pure and Applied Natural Sciences, University of Kalmar, SE-391 82 Kalmar, Sweden

**Gamma-ray bursts (GRBs) are the most powerful explosions in the universe and probe physics under extreme conditions. GRBs divide into two classes, of short and long duration[1], thought to originate from different types of progenitor systems[2,3]. The physics of their γ-ray emission is still poorly known, over 40 years after their discovery, but may be probed by their highest-energy photons. Here we report the first detection of high-energy emission from a short GRB with measured redshift, GRB 090510, using the Fermi Gamma-ray Space Telescope. We detect for the first time a GRB prompt spectrum with a significant deviation from the Band function. This can be interpreted as two distinct spectral components, which challenge the prevailing γ-ray emission mechanism: synchrotron – synchrotron self-Compton[4]. The detection of a 31 GeV photon during the first second sets the highest lower limit on a GRB outflow Lorentz factor, of >1200, suggesting that the outflows powering short GRBs are at least as highly relativistic as those powering long GRBs. Even more importantly, this photon sets limits on a possible linear energy dependence of the propagation speed of photons (Lorentz-invariance violation) requiring for the first time a quantum-gravity mass scale significantly above the Planck mass.**

On May 10th, 2009, $T_0$ = 00:22:59.97 UT, the Gamma-ray Burst Monitor (GBM)[5] and the Large Area Telescope (LAT)[6] on-board Fermi triggered on the very bright short GRB 090510, which was also detected and located with multiple satellites (Swift[7], Konus-Wind[8], Agile[9], Suzaku[10] and INTEGRAL-ACS). Ground-based optical spectroscopy data, taken 3.5 days later[11], exhibited prominent emission lines at a common redshift of $z = 0.903 \pm 0.003$, corresponding to a luminosity distance of $d_L = 1.8 \times 10^{28}$ cm (for a standard cosmology, $[\Omega_\Lambda, \Omega_M, h] = [0.73, 0.27, 0.71]$). The host galaxy of GRB090510 was identified as a late-type elliptical or early-type spiral star-forming galaxy, in contrast to the dwarf irregular, star-forming galaxies that have been observed to harbor long-duration GRBs[12,13], but consistent with the diverse types of hosts identified with short GRBs[14,15].

The GBM light curve (Fig. 1b,c; 8 keV – 40 MeV) consists of 7 main pulses. The main LAT emission above 100 MeV starts at $T_0$+0.65 s and lasts ~200 s; a single 31 GeV photon – the highest-energy photon detected from any GRB, coincides with the last GBM pulse at $T_0$+0.829 s (Fig. 1-b,c,f; see Supplementary Information 1). Using the combined high-energy resolution Time-Tagged Events from the three brightest GBM detectors we estimate the $T_{90}$ ($T_{50}$) duration of the event – the time within which 90% (50%) of the 50–300 keV event counts are collected – to be 2.1 s (0.2 s), placing GRB090510 within the short GRB range (see also Supplementary Information 2-A). We estimate the photon-energy associated delays (spectral-lags) of the emission, using two independent methods (described in Supplementary Information 2-B). We find no lags below 1 MeV (in agreement with the thus far known short GRB lags in that energy range), and above 30 MeV; however, we find that the bulk of the photons above 30 MeV arrive 258±34 ms later than those below 1 MeV.

We performed combined GBM+LAT spectral-fits from $T_0$ – 0.1 s to $T_0$ + 2.0 s (Table 1; see Supplementary Information 3). The *time-integrated* spectrum from $T_0$+0.5 s to $T_0$+1.0 s is best fit by two spectral components: a Band function (two power-laws smoothly joined near the peak photon-energy, $E_{peak}$) and a power-law. The addition of the latter component with a photon-index of −1.62±0.03 significantly improves the fit (>5 σ) compared to a single Band function and fits the data up to the highest-energy (31 GeV) photon (see Fig. 2a). We obtain $E_{peak}$ = 3.9±0.3 MeV - the highest Band function $E_{peak}$ ever measured in a GRB time-integrated spectrum. *Time-resolved* spectroscopy (Fig. 2c) from $T_0$ + 0.5 s to $T_0$ + 0.8 s, indicates that the GRB spectrum evolves from soft to hard (see also Table 1). The two spectral components (Band and power-law) show significant temporal correlation (while they are both detected; see supplementary information 3-A-3), suggesting that they originate from the same physical region (the 200 s extended LAT emission is not reported in this paper).

The progressively harder spectral behavior in GRB 090510 may be key to understanding particle acceleration and radiation in short GRBs. If the low-energy (10 keV – 10 MeV) spectral component is synchrotron radiation from non-thermal electrons, the high-energy (>10 MeV) power-law component may be synchrotron self-Compton[16,17]. However, the ~0.1–0.2 s observed delayed onset of the LAT high-energy photons is not expected in the simplest leptonic models unless magnetic fields, electron energy distribution or outflow Lorentz factors change rapidly during the GRB. A hadronic γ-ray emission component could arise from ultra-relativistic protons and ions, either through photo-meson[18] or synchrotron[19] processes. In the latter case, short GRBs would have to accelerate protons and ions to ultra-high energies, making short GRBs a possible source of ultra-high-energy cosmic rays, along with long GRBs and active galactic nuclei. This requires, however, a much larger total energy than that observed in γ-rays, thus straining the GRB energy-budget. The similarity in the high-energy emission of short and long GRBs, including delayed onset and temporally extended ~GeV radiation[20], suggests similarities in the underlying radiation physics of these two classes of GRBs, despite the currently prevailing paradigm that short GRBs originate from the coalescence of two neutron-stars or a neutron-star and a black-hole[2], whereas the long GRBs progenitors are very massive, rapidly rotating stars[3].

The distance of GRB 090510 is at the high-end of the redshift range for short GRBs and implies a very energetic event[21]. The total (0.5–1.0 s) energy fluence measured in the 10 keV – 30 GeV band is $(5.02 \pm 0.26) \times 10^{-5}$ ergs/cm$^2$, implying a total apparent isotropic energy release of $\approx (1.08 \pm 0.06) \times 10^{53}$ erg. The high-energy spectral component accounts for ~37% of the total fluence. The fluence could be higher if the emission extends to higher energies. However, the attenuation of the highest-energy photons by the extragalactic background light has a negligible ($\leq$ 1%) effect on the fluence in the observed energy-range (see Supplementary Information 4).

The spectra at the time intervals 0.6 s – 0.8 s and 0.8 s – 0.9 s include a 3.4 GeV and a 31 GeV photon, respectively. Such high-energy photons may pair-produce ($\gamma\gamma \rightarrow e^+e^-$) with radiation within the emitting region. To avoid such intrinsic attenuation, the emitting region must move toward the observer ultra-relativistically[22,23]. Thus for the 3.4 GeV and 31 GeV photons to pass through the lower-energy radiation within the source without pair-producing, the implied lower limits on the region bulk Lorentz factors, $\Gamma_{min}$, are 950±40 and 1220±60, respectively (see Supplementary Information 5 for details). The $\Gamma_{min}$ values we obtain for GRB090510 are the highest for any GRB[20], and by far the highest for a short GRB[21]. This suggests that the outflows powering short GRBs are at least as highly relativistic as those powering long GRBs.

Finally, the detection of multi-GeV photons <1 s from the onset of GRB 090510, and its known distance, allow us to tightly constrain Lorentz Invariance Violation (LIV). Some quantum-gravity theories[24] predict LIV, where the photon-propagation speed $v_{ph}$ depends on its energy $E_{ph}$ and is expected to significantly differ from the (low-energy) speed of light, $c \equiv v_{ph}(E_{ph}\rightarrow 0)$ only near the Planck scale ($E_{ph} \sim M_{Planck}c^2 \approx 1.22\times 10^{19}$ GeV)[24]. For $E_{ph} \ll M_{Planck}c^2$ the leading LIV term is expected to be $|v_{ph}/c - 1| \sim (E_{ph}/M_{QG,n}c^2)^n$ where $n = 1$ or 2 is usually assumed. For linear LIV ($n = 1$), this induces a difference $\Delta t = (\Delta E/M_{QG,1}c^2)D/c$ in the arrival time of photons emitted simultaneously at a distance $D$ from us, and differing by $\Delta E = E_{high} - E_{low}$ (at cosmological distances this simple expression is somewhat modified; see Supplementary Information 6-A). Because of their short duration and cosmological distances GRBs are well-suited for constraining LIV[24,25,20].

When allowing for LIV-induced time-delays, the measured arrival time, $t_h$, of the high-energy photons might not directly reflect their emission time, $t_{em}$ (i.e. their arrival time if $v_{ph} = c$). Therefore, we make reasonably conservative assumptions on $t_{em}$, based on the observed lower-energy emission (for which LIV-induced time-delays are relatively negligible). To constrain a positive time-delay ($v_{ph} < c$ implying $t_h > t_{em}$) we assume that $t_{em} > t_{start}$, where $t_{start}$ (see Fig. 1) corresponds to the onsets of the various lower-energy emission episodes at different photon-energies (see Supplementary Information 6 for discussion). This implies $\Delta t < t_h - t_{start}$ and thus sets a lower-limit on $M_{QG}$. For linear LIV ($n = 1$), the tightest limit can be placed by the photon with the highest $E_{high}/\Delta t$, which in GRB 090510, is the 31 GeV photon. We conservatively use the 1-σ lower-limits on the GRB redshift (z = 0.900) and photon energy (28.0 GeV; Table 2). Table 2 shows the resulting lower-limits on $M_{QG,1}$ for the different assumed values of $t_{start}$ for a positive time delay (see limits (a)–(d)), as well as time-delays of either sign, including negative ones ($v_{ph} > c$ implying $t_h < t_{em}$; limits (e) –(g)) ; these limits are all above $M_{Planck}$.

Our new limit, $M_{QG,1}/M_{Planck} \geq$ several (see Table 2), is much stronger than the previous best limit of this kind ($M_{QG,1}/M_{Planck} \geq 0.1$ from GRB080916C[20]) and fundamentally more meaningful. Since, in most quantum gravity scenarios, $M_{QG,n} \lesssim M_{Planck}$, even our most conservative limit (Table 2; (a)) greatly reduces the parameter space for n=1 models[26,27]. Our intermediate limits (Table 2; (b)–(d)), and even more so, our least conservative limit (Table 2; (e): $M_{QG,1}/M_{Planck} > 102$), based on associating the 31 GeV photon with the contemporaneous low-energy spike, makes such theories highly implausible (models with $n > 1$ are not significantly constrained by our results). Thus, we do not expect to see any evidence for other predictions of such $n = 1$ models, such as a reduction in the absorption of ≥10 TeV γ-rays by $\gamma\gamma \rightarrow e^+e^-$ interactions with extragalactic infrared photons[28] due to an increase in the required threshold energy, that would make the universe more transparent to γ-rays than expected[29,30].

**Acknowledgements** The Fermi LAT Collaboration acknowledges support from a number of agencies and institutes for both development and the operation of the LAT as well as scientific data analysis. These include NASA and DOE in the United States, CEA/Irfu and IN2P3/CNRS in France, ASI and INFN in Italy, MEXT, KEK, and JAXA in Japan, and the K. A. Wallenberg Foundation, the Swedish Research Council and the National Space Board in Sweden. Additional support from INAF in Italy for science analysis during the operations phase is also gratefully acknowledged. J.G. gratefully acknowledges a Royal Society Wolfson Research Merit Award. The Fermi GBM Collaboration acknowledges the support of NASA in the United States and DRL in Germany. We are grateful to John Ellis for helpful comments.

All authors contributed extensively to the work presented in this paper.

**Table 1 | Prompt Emission Spectral Fit Parameters**

| T– $T_0$ (s) | Model | Band | | | | PL or Comptonized | | | Castor |
|---|---|---|---|---|---|---|---|---|---|
| | | **A*$10^{-2}$** ph/cm$^2$/s/keV | **$E_{peak}$** MeV | $\alpha$ | $\beta$ | **A*$10^{-9}$** $\frac{\text{photons}}{\text{cm}^2\ \text{s keV}}$ at 1GeV | **$E_{peak}$** GeV | **Index** | |
| 0.5 – 1.0 | Band | $4.316^{+0.116}_{-0.115}$ | $4.104^{+0.267}_{-0.263}$ | $-0.75^{+0.03}_{-0.02}$ | $-2.40^{+0.04}_{-0.04}$ | | | | 1016/970 |
| | Band+PL | $3.188^{+0.269}_{-0.258}$ | $3.936^{+0.280}_{-0.260}$ | $-0.58^{+0.06}_{-0.05}$ | $-2.83^{+0.14}_{-0.20}$ | $2.426^{+0.531}_{-0.509}$ | | $-1.62^{+0.03}_{-0.03}$ | 979/968 |
| | Band +Comp | $3.203^{+0.281}_{-0.266}$ | $4.002^{+0.285}_{-0.271}$ | $-0.59^{+0.06}_{-0.06}$ | $-2.94^{+0.18}_{-0.25}$ | $3.011^{+0.697}_{-0.658}$ | $8.71^{+\infty}_{-4.18}$ | $-1.60^{+0.03}_{-0.03}$ | 976/967 |
| 0.5 – 0.6 | Band | $8.047^{+0.346}_{-0.344}$ | $2.809^{+0.185}_{-0.174}$ | $-0.59^{+0.04}_{-0.04}$ | < –5.0 | | | | 840/971 |
| 0.6 – 0.8 | Band+PL | $2.984^{+0.365}_{-0.341}$ | $5.102^{+0.443}_{-0.400}$ | $-0.48^{+0.07}_{-0.07}$ | $-3.09^{+0.21}_{-0.35}$ | $1.862^{+0.719}_{-0.625}$ | | $-1.66^{+0.04}_{-0.04}$ | 991/968 |
| 0.8 – 0.9 | Band | $0.040^{+0.005}_{-0.004}$ | $1.414^{+0.928}_{-0.536}$ | $-1.00^{+0.11}_{-0.09}$ | $-1.85^{+0.05}_{-0.06}$ | | | | 886/970 |
| | Band+PL | $0.028^{+0.006}_{-0.006}$ | $1.894^{+1.160}_{-0.718}$ | $-0.86^{+0.17}_{-0.23}$ | $-3.09$ (fixed) | $6.439^{+1.550}_{-1.230}$ | | $-1.54^{+0.07}_{-0.04}$ | 890/969 |
| 0.9 – 1.0 | PL (LAT only) | | | | | $3.721^{+1.260}_{-1.080}$ | | $-1.92^{+0.20}_{-0.22}$ | 43/118 |

**Upper section**: fits of the time-integrated spectrum of GRB090510 from $T_0$+0.5 s to $T_0$+1.0 s. The Band+PL model improves significantly (>5 $\sigma$) the fit compared to the standard Band function. The Band+Comptonized fit implies a lower-limit of 4 GeV for the energy of a possible spectral-break (softening) in the additional power-law. The various models are compared using the castor Castor statistic (C-stat), which is a likelihood method similar to $\chi^2$ when there is high enough statistics. Simple comparison of the C-stat values and degrees of freedom between the various models allows us determine the best model.

**Lower section**: best-fit spectral-models of GBM+LAT data for each selected time-interval from $T_0$+0.5 s to $T_0$+1.0 s, corresponding to the main GBM emission episode. The spectrum during 0.6 – 0.8 s shows a clear excess relative to a single Band function at high energies, which is best fitted with an additional power-law, indicating an additional spectral component. At 0.8 – 0.9 s, the additional component is not statistically justified anymore and both a Band function and Band+PL provide a good fit, if we fix β to its value from the previous time bin. At later times, a single Band function is marginally preferred over a Band+PL fit.

**Table 2 | Limits on Lorentz Invariance Violation**

| # | $t_{start} - T_0$ (ms) | Limit on $\|\Delta t\|$ (ms) | Reasoning for choice of $t_{start}$ or limit on $\Delta t$ or $\|\Delta t/\Delta E\|$ | $E_l$ † (MeV) | Valid for $s_n$* | Lower limit on $M_{QG,1}/M_{Planck}$ |
|---|---|---|---|---|---|---|
| (a)❖ | −30 | < 859 | start of any < 1 MeV emission | 0.1 | 1 | > 1.19 |
| (b)❖ | 530 | < 299 | start of main < 1 MeV emission | 0.1 | 1 | > 3.42 |
| (c)❖ | *648* | < 181 | start of main > 0.1 GeV emission | 100 | 1 | > 5.63 |
| (d)❖ | 730 | < 99 | start of > 1 GeV emission | 1000 | 1 | > 10.0 |
| (e)♦ | — | < 10 | association with < 1 MeV spike | 0.1 | ±1 | > 102 |
| (f)♦ | — | < 19 | If 0.75 GeV‡ γ-ray from 1st spike | 0.1 | −1 | > 1.33 |
| (g)♠ | $\|\Delta t/\Delta E\|$ < 30 ms/GeV | | lag analysis of > 1 GeV spikes | — | ±1 | > 1.22 |

Details for the derivations of these limits are given in the Supplementary Information. Limits (a) through (e) rely on the 31 GeV photon, and use the 1-σ lower limit on its energy (28.0 GeV). The 1-σ lower limit on the redshift (z = 0.900) is used for all our limits.

†The typical energy of the low-energy photons that were used for reference.

*$s_n$ = 1 and -1 stand for a positive ($v_{ph}$ < c) and negative ($v_{ph}$ > c) time-delay, respectively.

‡We conservatively used the 1-σ lower limit on this photon's energy (0.694 GeV).

❖The different choices of $t_{start}$ in limits (a)–(d) are ordered from the most conservative, (a), to the somewhat less conservative but still very reasonable, (d); for a detailed discussion see the Supplementary Information 6.

♦We constrain time-delays of either sign, including negative ones ($v_{ph}$ > c implying $t_h < t_{em}$) using two methods; (e)–(f) by associating two high-energy photons with contemporaneous spikes in the low-energy lightcurve (see also the gray shaded regions in Fig. 1).

♠(g) by testing for an energy-dispersion in the high-energy lightcurves (that might smear the sharp observed spikes) we find an upper limit for a linear dispersion of photons above 100 MeV of $|\Delta t/\Delta E|$ < 30 ms/GeV (at 99% confidence; see Supplementary information 2-B-2).

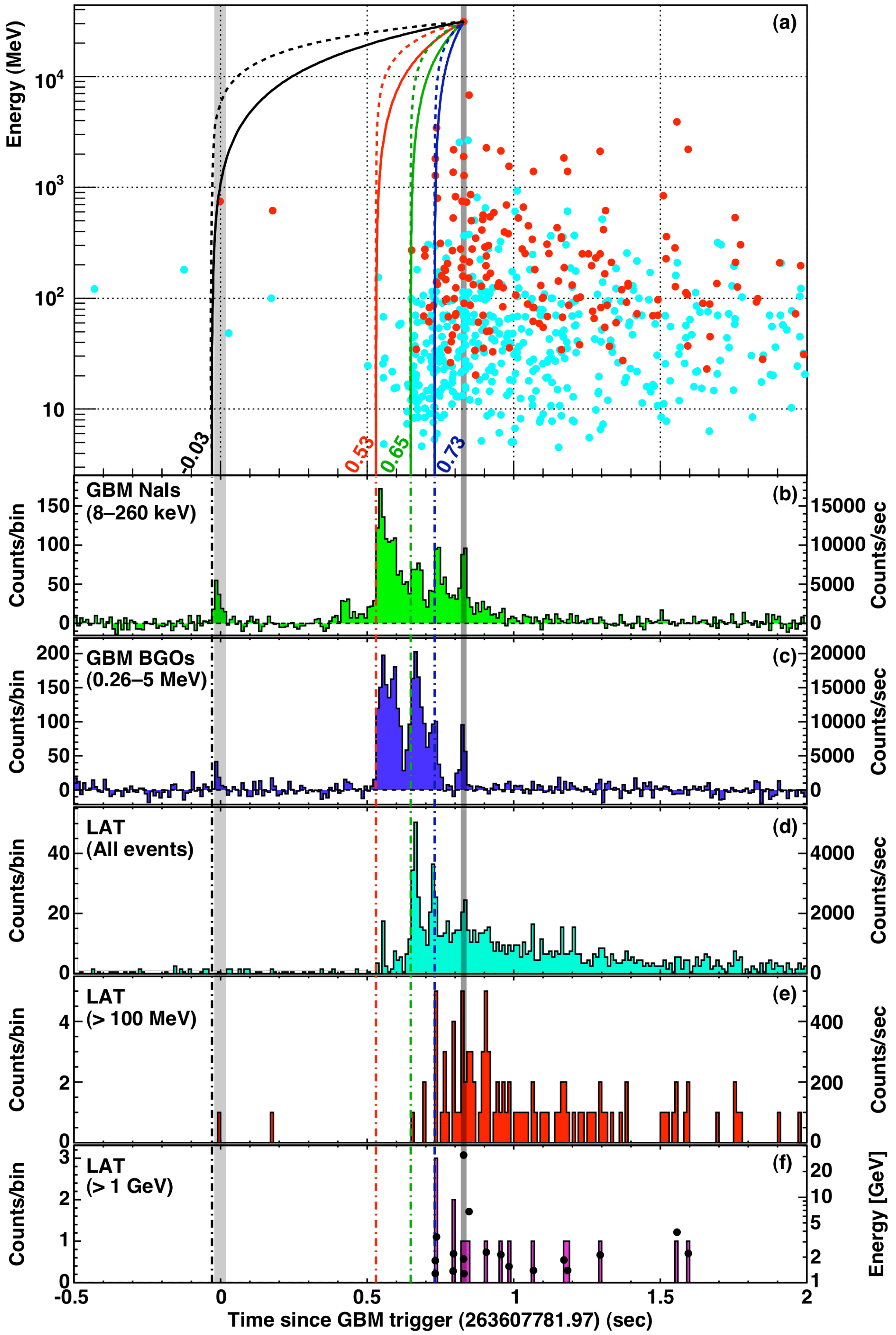

**Figure 1. Panel (a)**: energy vs. arrival time w.r.t the GBM trigger time for the 160 LAT photons that passed the transient off-line event selection (*red*) and the 161 photons that passed the onboard $\gamma$-ray filter (*cyan*), and are consistent with the direction of GRB 090510. The *solid* and *dashed* curves are normalized to pass through the highest energy (31 GeV) photon and represent the relation between a photon's energy and arrival time for linear ($n$=1) and quadratic ($n$=2) LIV, respectively, assuming it is emitted at $t_{start} - T_0 = -30$ ms (*black*; first small GBM pulse onset), 530 ms (*red*; main <MeV emission onset), 648 ms (*green*; >100 MeV emission onset), 730 ms (*blue*; >GeV emission onset). Photons emitted at $t_{start}$ would be located along such a line due to (a positive) LIV induced time delay. **Panels (b)–(f):** GBM and LAT lightcurves, from lowest to highest energies. Panel (f) also overlays energy vs. arrival time for each photon, with the energy scale displayed on the right side. The *dashed-dotted* vertical lines show our 4 different possible choices for $t_{start}$. The gray shaded regions indicate the arrival time of the 31 GeV photon ±10 ms (*on the right*) and of a 750 MeV photon (*during the first GBM pulse*) ±20 ms (*on the left*), which can both constrain a negative time delay. Panels (b) and (c) show background subtracted lightcurves for GBM NaI in the 8–260 keV band and a GBM BGO in the 0.260–5 MeV band, respectively. **Panels (d)–(f)** show, respectively, LAT events passing the onboard $\gamma$-ray filter, LAT transient class events with E > 100 MeV, and LAT transient class events with E > 1 GeV. In all lightcurves, the time-bin width is 10 ms. In panels (b)–(e) the per-second count rate is displayed on the right for convenience.

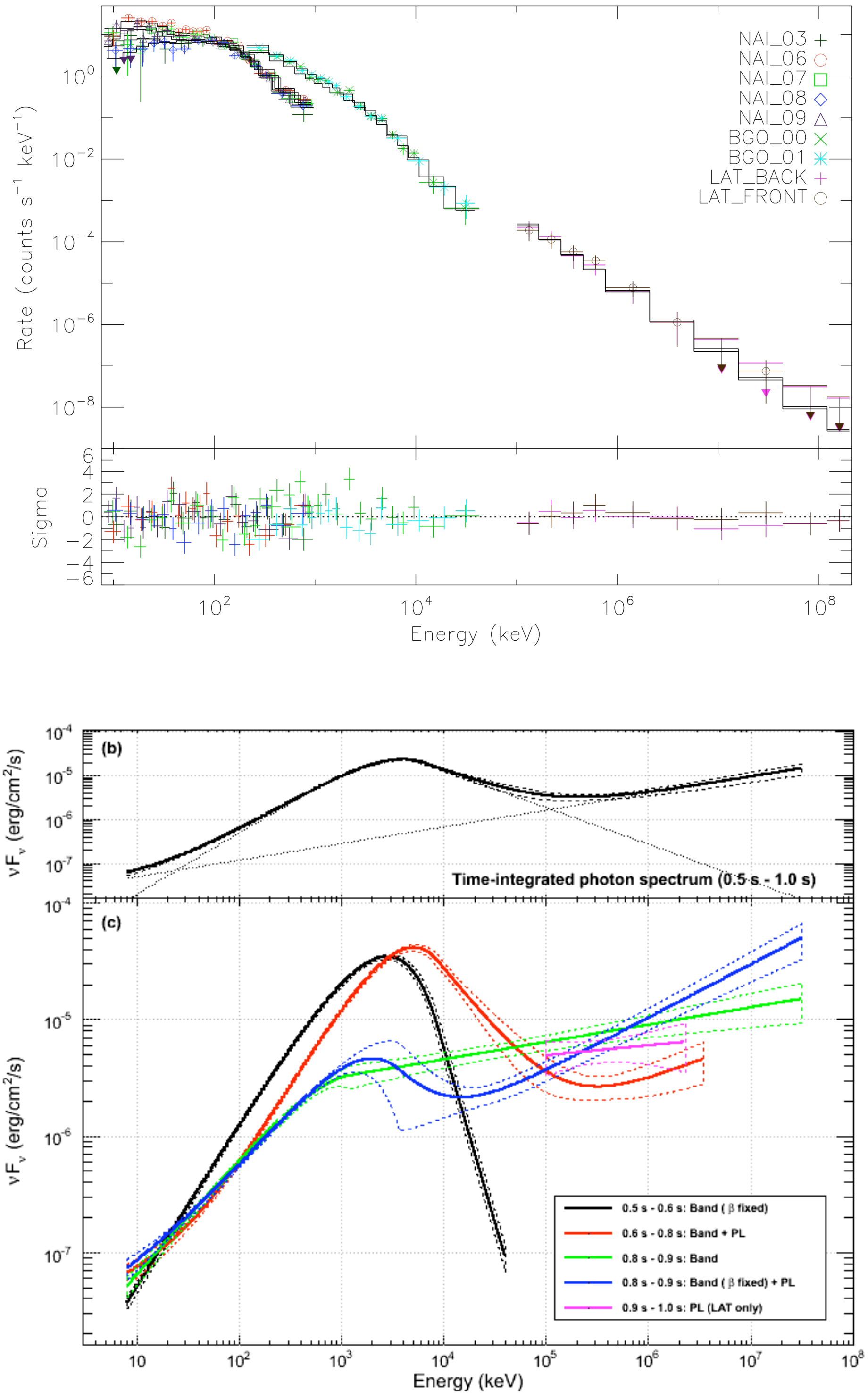

**Figure 2. Panel (a)**: time-integrated count-spectrum from $T_0$+0.5 s to $T_0$+1.0 s. Following a ~0.5 s gap after the first GBM small peak, comes the most intense part of the burst, comprising 5 pulses, lasting from ~$T_0$+0.5 s to ~$T_0$+0.9 s. The peak of the emission (at $T_0$+0.54 s), appears during the second pulse, which is detected up to ~10

MeV with significant substructure. The third pulse however, although weaker, shows up to much higher energies (~40 MeV) and is visible in GBM and in the LAT all-events light curve, showing a delayed onset of the high-energy emission. The overall LAT emission is visible for ~ 200 s. Front and back LAT events are analyzed as if from separate detectors; the signal is highly significant up to ~5 GeV. The data are adequately fit by a Band+PL spectrum as indicated by the fit-residuals. **Panel (b)**: best-fit spectral model (Band+PL) for the time-integrated (0.5 – 1.0 s) spectrum (see also text and upper part of Table 1). **Panel (c)**: spectral evolution from $T_0$+0.5 s to $T_0$+1.0 s (see also lower part of Table 1). While a single component (Band function with a very soft $\beta$) is adequate during $T_0$+(0.5 – 0.6) s, an additional power-law component is required only at $T_0$+(0.6 – 0.8) s. After $T_0$+0.9 s the signal in GBM is too weak and the LAT only data can be fit by a power-law. However, this power-law is inconsistent with the GBM upper-limits, requiring a spectral break between the LAT and GBM energy-range